\documentclass{elsart}

\usepackage{harvard}

\usepackage{graphicx}

\usepackage{amssymb}

\def\url#1{{\ttfamily\def\/{/\discretionary{}{}{}}#1}}

\begin{document}

\begin{frontmatter}
\title{Restarting activity in radio galaxies}

\author[IAA]{L. Lara}, 
\author[BOL,IRA]{G. Giovannini},
\author[NRAO1]{W.D. Cotton},
\author[IRA]{L. Feretti},
\author[UVAL]{J.M. Marcaide},
\author[IAA]{I. M\'arquez},
\author[NRAO2]{G.B. Taylor},
\author[IRA]{T. Venturi}

\address[IAA]{Instituto de Astrof\'{\i}sica de Andaluc\'{\i}a, CSIC, Apdo. 
3004, 18080 Granada, Spain} 
\address[BOL]{Dpto. di Fisica, Universit\'a di Bologna, Via Pichat 6/2, 
40127 Bologna, Italy}
\address[IRA]{Istituto di Radioastronomia, Via P. Gobetti 101, 40129 Bologna, 
Italy}
\address[NRAO1]{NRAO, 520 Edgemont Road, Charlottesville, VA 22903-2475, USA}
\address[UVAL]{Dpto. de Astronom\'{\i}a, Universitat de Val\`encia, 46100 
Burjassot, Valencia, Spain}
\address[NRAO2]{NRAO, P.O. Box O, Socorro, NM 87801, USA}

\begin{abstract}
We present observations of two radio galaxies, J1835+620 and 3C338, 
both with signs of having passed through different stages of core 
activity. The former presents two symmetric and bright components within 
a typical FR II structure, possibly resulting from two distinct phases 
of activity; the latter is a FR I radio galaxy with two separated 
regions with different age properties, possibly due to a switch-off and 
-on cycle in its core. In both sources, the optical counterpart lies 
in a group of galaxies with indications of mutual interaction, a scenario 
often invoked to explain triggering of core activity. 
\end{abstract}

\end{frontmatter}

\section{Introduction}
\label{intro}

The  large diversity of  radio  sources, partially understood as 
the result of orientation related effects \cite{urry},  seems to suggest 
also an evolutionary process in the lifetime of a radio source.  There is  
increasing evidence that  Compact Symmetric Objects are young 
radio sources; that typical FR I and FR II radio galaxies may  be 
considered ``adult''  sources; and finally, that many  of the 
relic  sources are  probably  associated with  dormant radio galaxies 
which have ceased their nuclear activity. If we believe  that the activity 
in radio-loud  AGNs is the  result of  accretion onto a  compact massive 
object,  likely a 
black hole,  the life of a  radio source would be controlled by the 
accretion rate.  Under  such a scenario, it should  be expected that a 
significant  number of radio  sources with  clear evidences  of having 
passed through  different phases during  their lifetime are  found. We 
present here observations of two such radio galaxies, J1835+620 and 3C338.

\section{The Giant Radio Galaxy J1835+620}
\label{1835}

\begin{figure}
\begin{center}
\includegraphics*[scale=0.6,angle=0]{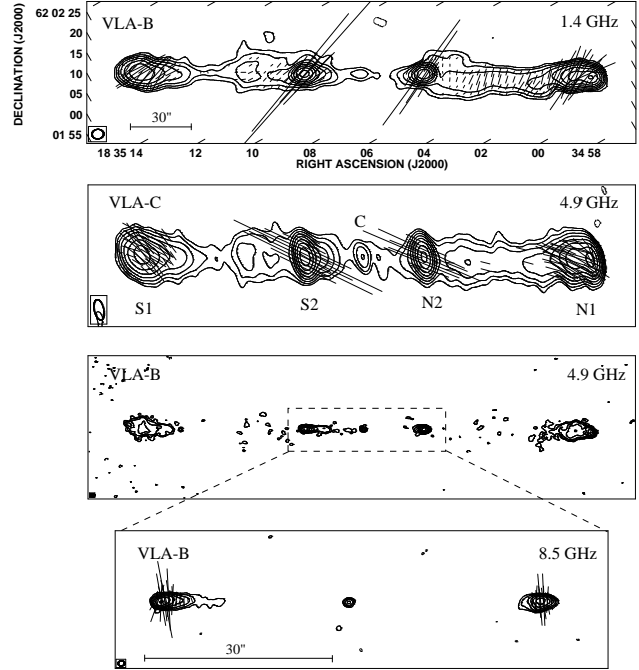}
\end{center}
\caption{VLA maps of J1835+620 at 1.4, 4.9 and 8.5 GHz, 
rotated clockwise on  the sky  by $60^{\circ}$.    Contours are spaced   by 
factors of 2 in brightness,  with the lowest at 3  times the rms noise 
level.   The vectors    represent  the  polarization   position  angle 
(E-vector),   with   length     proportional  to   the    amount    of 
polarization. $1''$ corresponds to 4.81 Kpc.} 
\label{fig1}
\end{figure}

\begin{figure}
\begin{center}
\includegraphics*[scale=0.7,angle=0]{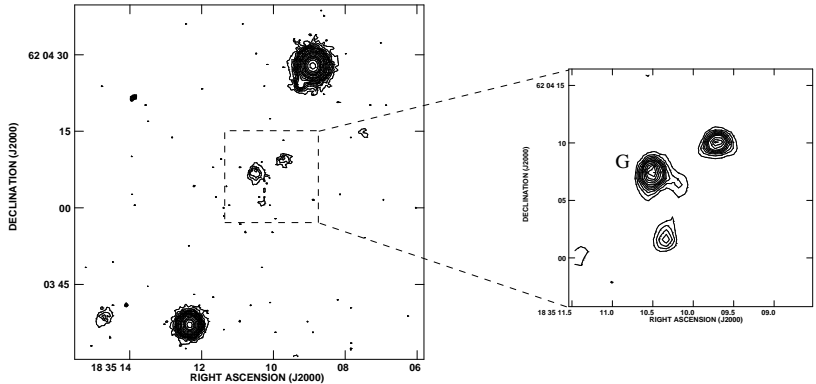}
\end{center}
\caption{Optical observations of J1835+620: The left panel shows 
an R-filter image  centered at the  position of the  radio core.   The
right  panel  shows  a detail  of the  optical  image after  the
application of deconvolution algorithms.  The host galaxy of J1835+620
is  labeled with  the letter ``G''.}
\label{fig2}
\end{figure}

The radio source J1835+620 belongs to a new sample of large angular size 
radio galaxies (Lara et  al., in preparation) selected  from the NRAO  VLA 
Sky Survey \cite{condon}.  We have observed this source  with the VLA in 
its B-  and C- configurations  at 1.4, 4.9  and 8.5 GHz, and  with the 
MPIfA 2.2m optical telescope in Calar Alto (Spain). The  outstanding  aspect 
of  J1835+620  at  radio  wavelengths is  the 
existence  of two  symmetric bright  components (N2  and S2)  within a 
typical FR II structure  \cite{lara,arno}(see {Fig.~\ref{fig1}}). Its 
optical counterpart, coincident with the radio core (C) has a spectrum 
with strong  and narrow emission lines. It lies in a group  of at  least 
three (probably four) galaxies showing signs  of mutual interaction 
(Fig.~\ref{fig2}). With an angular size of 3.88' and a redshift of z=0.518, 
implying a total source length of 1.12 Mpc (H$_0$= 75 km s$^{-1}$ Mpc$^{-1}$; q$_o$=0.5), J1835+620 belongs to  the group of giant radio galaxies.

We can interpret the peculiar structure of J1835+620 as the result of two 
distinct phases of core activity. However, the existence  of a hot-spot  
in component  N1  together with  spectral  aging  arguments indicate  that  
N1 and  S1  are  still  supplied by  fresh  particles, implying {\em  i)} 
the existence  of an underlying jet  connecting the core  with the outer  
components and  {\em ii)}  that the  activity in J1835+620 did not stop  
completely. In consequence, the new components N2 and  S2 would  represent 
the result  of a new  ejection propagating through the primary underlying 
jet.  The parallel configuration of the magnetic field in  components N2 
and S2 (Fig.~\ref{fig1}, bottom panel) is  consistent with  a 
``second-phase''  jet which  is  overdense with respect to the 
``first-phase'' one \cite{lara}.

\section{The FR I Radio Galaxy 3C338}
\label{3C338}

3C338 ($z=0.030$) is  a FR I radio source,  associated with the multiple
nuclei cD galaxy NGC 6166 at the center of the cooling flow cluster of
galaxies  A2199.   We  are  undertaking  a  program  of  VLA  and  VLBI
observations  of 3C338  in  order  to study  its  strong flux  density
variability, its large scale  properties and the structural variations
at parsec scales \cite{giov}.

\begin{figure}
\begin{center}
\includegraphics[scale=0.5,angle=0]{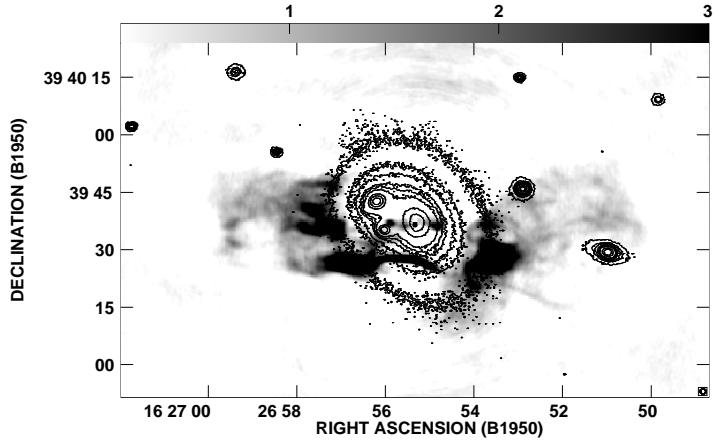}
\end{center}
\caption{In greyscale, VLA map of 3C338 at 1.7 GHz. 
It was 
deconvolved with the maximum entropy method. The nuclear source has been 
partially subtracted. Contours represent an optical CCD image, courtesy of 
G. Gavazzi.}
\label{fig3}
\end{figure}

The large-scale structure of 3C338 can  be separated in  two regions with
very different properties: an  active region, which includes the core,
two symmetric  jets and  two faint hot  spots at  the jet ends,  and a
diffuse region, displaced to the south, showing a jetlike filament and
low-brightness  extended  emission  \cite{giov} (Fig.~\ref{fig3}).  If  the 
radio core  stopped or decreased its  activity at
some  time,  it  could  have  left  a
steep-spectrum  aged radio jet  behind (relic emission).  
Burns  et al.
(1983) suggested two  possible explanations for the existence of two
separated regions in this peculiar source:
(1) the ram pressure  of a highly asymmetric cooling  flow onto the cD
galaxy or (2) the motion of the radio core within the cD galaxy.
Alternatively, the relic emission could have been produced by past activity in one of the other nuclei in the cD galaxy. However, the relic and the new emission being quite parallel and the curvature of the relic filament in agreement with Burns et al. suggestions  make this interpretation less attractive. 

\begin{figure}
\begin{center}
\includegraphics[scale=0.5,angle=-90]{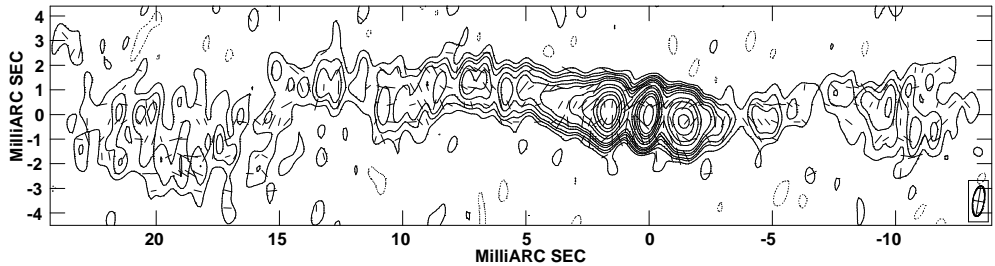}
\end{center}
\caption{VLBI map of 3C338 made at 8.4 GHz in September 1997. (Giovannini et al., in preparation)}
\label{fig4}
\end{figure}

The radio structure of the small region is similar to that of extended
FR  II radio  galaxies but  on a  much smaller  scale, similar  to the
high-power  medium-sized  symmetric  objects  found at  high  redshift
\cite{readhead}. At parsec  scales, 3C338 has an
unusual structure consisting of a compact core and two symmetric jets,
with moving components in both directions (Fig.~\ref{fig4}).

\section{Conclusions}

We present  observational results on two very different radio  galaxies, 
the giant  FR II J1835+620 and the  FR I  3C338, both with  radio 
structures showing  evidence of distinct  phases of  activity  during 
their  lifetimes. 
If  the  activity is  intimately
related  to  the  accretion  of  matter onto  a  massive  object,  the
vanishing or diminution of accretion  would lead a former radio source
to a ``dormant'' or hibernation  phase. Such a picture is supported by
the  increasing number of  massive dark  objects detected  in inactive
galaxies \cite{kormendy}. Interaction and
merging  with  neighboring  galaxies  can trigger  the  activity,  and
eventually  produce a  transition from  a dormant  to an  active-core phase
\cite{stockton,bahcall}. We note that J1835+620  and 3C338  have both optical  counterparts  with 
nearby  companions,  with signs of  mutual interaction (Figs.~\ref{fig2}, 
\ref{fig3}), supporting  
the previous scenario.

\end{document}